\documentclass[twocolumn,nofootinbib,preprintnumbers,amsmath,amssymb]{revtex4}

\usepackage{graphicx}
\usepackage{dcolumn}
\usepackage{bm}
\usepackage{amsmath}

\def \be {\begin{equation}}
\def \ee {\end{equation}}

\begin{document}

\title{Pulsar Emission Spectrum}

\author{Andrei Gruzinov}

\affiliation{ CCPP, Physics Department, New York University, 4 Washington Place, New York, NY 10003
}

\begin{abstract}

Emission spectrum is calculated for a weak axisymmetric pulsar. Also calculated are the observed spectrum, efficiency, and the observed efficiency. The underlying flow of electrons and positrons turns out to be curiously intricate.

\end{abstract}

\maketitle

\section{Introduction}

Here we calculate pulsar emission, apparently from first principles, i.e., without any arbitrary assumptions. Our method applies to weak pulsars.  A pulsar is called weak if pair production near the light cylinder is negligible compared to Goldreich-Julian density \cite{Goldreich1969} per rotation. The weak pulsar approximation must be accurate for some, and perhaps meaningful for most, Fermi \cite{Fermi2013} pulsars. 

We treat only the axisymmetric case. But the method is readily applicable in 3D and is expected, in view of the encouraging axisymmetric results, to yield energy-resolved light curves which will agree with observations without any adjustable parameters (except the inevitable two: spin-dipole and observation angles). 

The following almost trivial observation appears to be crucial for calculating pulsars, both weak and strong. For pulsar-strength electromagnetic fields the Lorentz equation for the {\it acceleration} of charges should be replaced by an ``Aristotelian'' expression for the {\it velocity} of charges, Eq.(1). In combination with Maxwell equations, we will call this Aristotelian Electrodynamics (AE).

AE can already claim some tentative successes in explaining pulsars. Firstly, for weak pulsars, AE predicts the averaged efficiency of order 10\% \cite{Gruzinov2012} -- right in the middle of what's seen by Fermi \footnote{ As shown below, weak pulsars should be calculated with zero multiplicity in the radiation zone. We have re-done the simulation of \cite{Gruzinov2012} accordingly and seen no significant changes in efficiency. But our 3D code has a large star,  small box, and  poor resolution. We find a factor of about 1.6 greater than \cite{Gruzinov2012} axisymmetric efficiency in \S\ref{emiti}. }. Secondly, AE correctly predicts the characteristic photon energy for weak pulsars \cite{Gruzinov2012}. Thirdly, AE explains the decreased efficiency of strong pulsars \cite{Gruzinov2013}. Here we show that AE also gives reasonable observed spectra and observed efficiencies for weak pulsars (``observed'' means emitted into a fixed infinitesimal solid angle).

In AE charges move with prescribed velocities rather than accelerations. Positrons and electrons move, as we show in \S\ref{deri}, at the speed of light (see \S\ref{deri} for a clarification):
\be\label{vel}
    {\bf v}_{\pm}={{\bf E}\times {\bf B}\pm(B_0{\bf B}+E_0{\bf E})\over B^2+E_0^2}.
\ee
Here $E_0$ is the proper electric field scalar and $B_0$ is the proper magnetic field pseudoscalar defined by 
\be\label{prop}
B_0^2-E_0^2=B^2-E^2,~ B_0E_0={\bf B}\cdot {\bf E},~ E_0\geq 0.
\ee

Knowing how the charges move, and calculating or postulating the pair production rate, we can calculate the electric current and solve Maxwell equations. This has been done in \cite{Gruzinov2013}, where the charges have been represented by the density fields. Unfortunately, the numerical scheme of \cite{Gruzinov2013} seems to be somewhat flawed. Apparently due to interpolations in the force-free zone, where the sign of $B_0$ fluctuates, one gets particle velocities which are not aligned with Eq.(\ref{vel}). 

To remedy this flaw, we performed a particle simulation of AE, where individual charges are described by their coordinates. To the extent that it worked, the particle simulation seems to agree with the density simulation \cite{Gruzinov2013}, and by construction it does not give wrongly aligned currents. But the accuracy and stability of our particle simulations are poor.

It appears however, and quite unexpectedly, that the weak pulsar problem can be solved by a simplified procedure which avoids explicit calculation of the densities of electrons and positrons. We calculate only the electromagnetic field using a certain Ohm's law to close the system of Maxwell equations. The resulting magnetosphere turns out to be realizable as an AE flow emanating from the star. Moreover, the magnetosphere has some special properties which allow to infer the densities of electrons and positrons (not everywhere, but everywhere where it matters) and then calculate the emission spectrum.

Still we must label our results as tentative until confirmed by a full AE simulation. Even though the calculated magnetosphere is similar to what we got in \cite{Gruzinov2013} (without the non-alignment issues of \cite{Gruzinov2013}) and similar to what we got in a particle simulation (without the bad accuracy and stability issues), it is possible that multiple AE flows exist, and the true (occurring in nature) AE flow is not what our indirect procedure finds; or the true AE flow can be turbulent.

In \S\S\ref{magnetos},\ref{emiti} we calculate the magnetosphere and radiation. For clarity, these sections are written in a recipe form; explanations and justifications are in \S\ref{deri}.

\section{Magnetosphere}\label{magnetos}

First solve Maxwell equations
\begin{eqnarray}\label{max}
\dot{{\bf B}}=-\nabla \times {\bf E},
\\
\dot{{\bf E}}=\nabla \times {\bf B}-{\bf j},
\end{eqnarray}
in entire space. The current ${\bf j}$ is given by two different expressions, which we call Ohm's laws, inside and outside the star of radius $R_s$.

\begin{figure}[bth]
  \centering
  \includegraphics[width=0.48\textwidth]{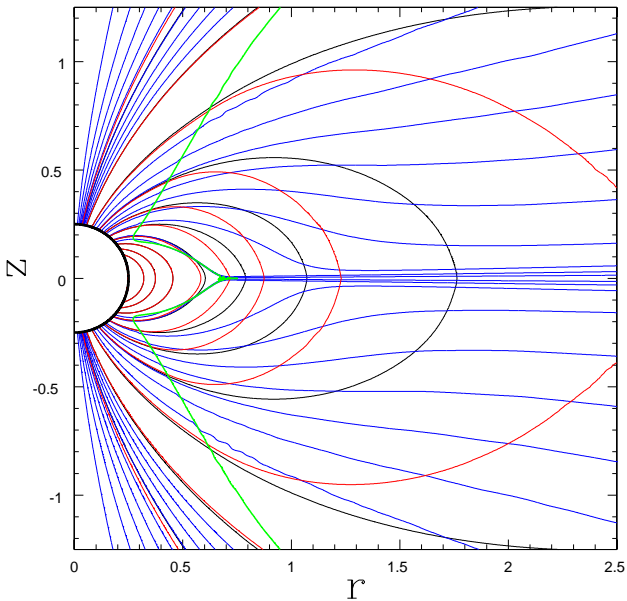}
\caption{Thin blue: $A$, integer multiples of $0.1A_{\rm max}$, $A_{\rm max}=0.25$. Black and thin red: $\psi$ and $\phi$, 0.1,0.2, and integer multiples of 0.4. Thick green: the radiation zone boundary.} \label{magn}
\end{figure}

Inside the star, for spherical radii $R<R_s$, use the standard Ohm's law (in a rotating frame, angular velocity $\Omega$) plus a fixed external toroidal current. The conductivity of the star is arbitrary, but $\gg 1/R_s$. The value and poloidal profile of the external current are arbitrary; they just determines the magnetic moment of the star $\mu$; and $\mu$ scales out of the final results which will be given in terms of the pulsar observables: the spin-down power $L_{\rm sd}$ and the angular velocity  of the star $\Omega$. 

Outside the star,  $R>R_s$, use the following Ohm's law
\be\label{ohm}
    {\bf j}={\rho {\bf E}\times {\bf B}+|\rho |(B_0{\bf B}+E_0{\bf E})\over B^2+E_0^2},
\ee
where 
\be
\rho \equiv \nabla \cdot {\bf E}
\ee
is the charge density.

Start with initial fields equal to zero. Regularize Maxwell equations (\ref{max}) by small diffusivities $\eta \nabla^2{\bf B}$ and $\eta \nabla^2{\bf E}$ with $\eta \ll R_s$. Regularize the Ohm's law (\ref{ohm}) by $E_0\rightarrow E_0+\epsilon$ with $\epsilon \ll$ the characteristic magnetic field value. 

After a while the electromagnetic field saturates at what is shown in Fig.\ref{magn}. All results (unless specified otherwise) are given in ``pulsar units'':
\be
c=\Omega=R_{\rm lc}=\mu=1,
\ee
where $R_{\rm lc}=c/\Omega$ is the light cylinder radius.

In cylindrical coordinates, $(r,\theta,z)$, the saturated fields depend only on $r$ and $z$ and can be written as 
\begin{eqnarray}\label{pot}
{\bf E}=\left( -\partial _r\phi, 0, -\partial _z\phi \right) ,
\\
{\bf B}={1\over r}\left( -\partial _z\psi, A, \partial _r\psi \right) .
\end{eqnarray} 
The electric field is represented by the electrostatic potential $\phi$. The poloidal magnetic field is represented by the ``magnetic stream function'' $\psi$,  equal to the $\theta$-component of vector potential divided by $r$. The toroidal magnetic field is represented by the quantity $A$ equal to twice the integrated poloidal current.

The thick green line in Fig.\ref{magn} is the boundary between the force-free zone and the radiation zone. In the force-free zone, which lies inside the green line, $E_0=0$. In the radiation zone, which lies outside the green line, $E_0>0$.

The electric current flows along magnetic surfaces $A=const$. It is seen that the current flows everywhere in the radiation zone, while part of the force-free zone is free from poloidal current: this is the corotation zone, where the charges just rotate with the angular velocity of the star $\Omega$. 

Calculating the Poynting flux emanating from the star, $L_{\rm sd}={1\over 2}\oint d\phi A$, we get the spin-down power 
\be
L_{\rm sd}\approx 0.3{\mu ^2\Omega^4\over c^3},
\ee
which is well below the standard force-free pulsar luminosity\footnote{ For the force-free pulsar (first calculated by \cite{Contopoulos1999}, improved by \cite{Gruzinov2005}, done by \cite{Spitkovsky2006}), $L_{\rm sd}\approx {\mu ^2\Omega^4\over c^3}(1+\sin ^2\theta)$, where $\theta$ is the spin-dipole angle. But as discussed at the end of \S3.4 of \cite{Gruzinov2013}, the very existence of force-free pulsars is questionable.}. 

We must state that our numerical simulation has insufficient resolution (we use the same primitive code as in \cite{Gruzinov2013} except that now we calculate the electric current from the Ohm's law instead of particle densities). It appears that one needs a star smaller than our $R_s=0.25$ and a simulation box larger than our 5x10. As it is, our simulation results do not quite converge even at our ultimate 1600x3200 resolution. To take a crude guess, our numerical values are perhaps $\sim$10\% uncertain. 

\section{Emission}\label{emiti}

Given the electromagnetic field, one calculates emission as follows. The force-free zone does not radiate. In the radiation zone, the radiated power per unit volume is
\be
q=c|\rho|E_0 .
\ee
This power is emitted along 
\be 
{\bf v} = \left\{ \begin{array}{rl}
{\bf v}_+, & \rho >0; \\
{\bf v}_-, & \rho <0;
\end{array}\right.
\ee
where ${\bf v}_\pm$ are from eq.(\ref{vel}). The power is emitted with synchrotron spectrum of critical photon energy
\be\label{ecrit}
E_{\rm c}=(3/2)^{7/4}c\hbar e^{-3/4}E_0^{3/4}K^{-1/2},
\ee
where the curvature $K$ is given by
\be
K=|{\bf v}\cdot \nabla {\bf v}|.
\ee

\begin{figure}[bth]
  \centering
  \includegraphics[width=0.48\textwidth]{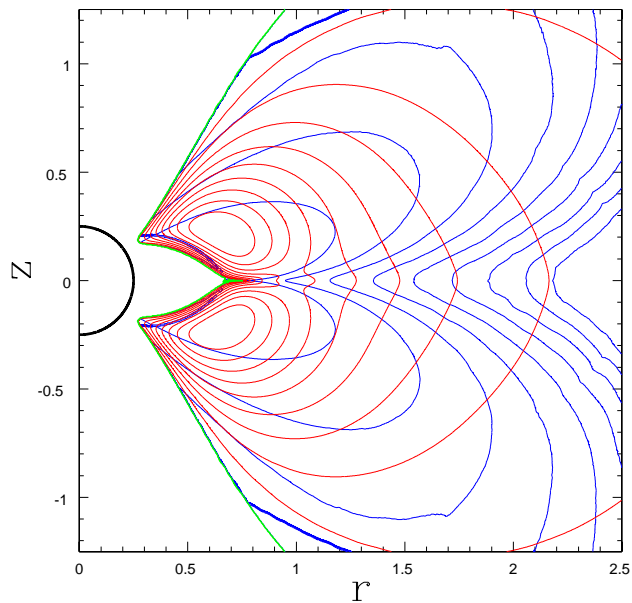}
\caption{Thin blue: radius of curvature to cylindrical radius ratio, $(Kr)^{-1}$, integer multiples of 0.5, increasing with increasing $r$. Red: $E_0$, integer multiples of $0.1E_{0~{\rm max}}$, $E_{0~{\rm max}}=0.65$. } \label{curv}
\end{figure}

\begin{figure}[bth]
  \centering
  \includegraphics[width=0.48\textwidth]{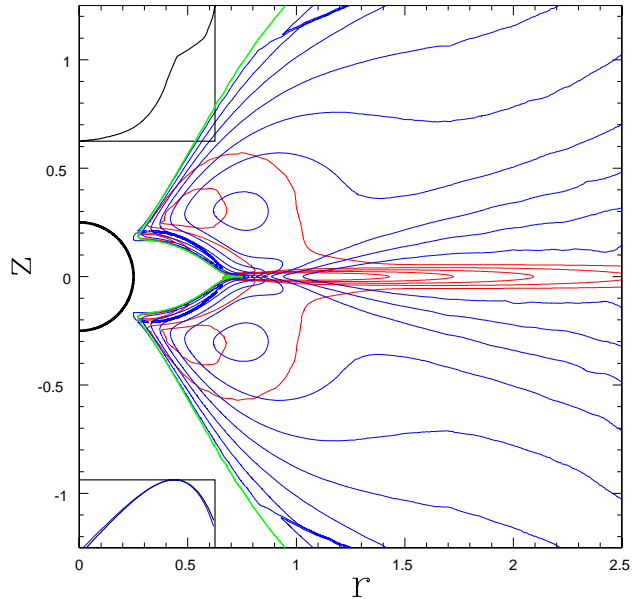}
\caption{Thin blue: critical energy of the photon spectrum $E_{\rm c}$, integer multiples of $0.1E_{\rm c~max}$ with $E_{\rm c~max}=2.5{mc^2\over \alpha}{\rm Ar}^{3/8}$. Red: $r$-normalized power density $rq$, in octaves, from 0.05 to 3.2. Upper inset: power distribution over critical energies -- total power emitted at critical energies $<E_{\rm c}$ vs. $E_{\rm c}$. Lower inset: power spectrum, $\log (E^2{dN\over dE})$ vs. $\log E$, and the PLEC fit (thin blue). } \label{ecr}
\end{figure}

The proper electric field and the curvature are shown in Fig.(\ref{curv}). The radiated power density and the critical photon energy as a function of position are shown in Fig.(\ref{ecr}). The insets show the power emitted at critical energies less than $E_{\rm c}$ as a function of $E_{\rm c}$ and the corresponding emission spectrum (obtained by convolving the spectrum of critical energies with the  synchrotron spectrum); also shown is the PLEC fit (power law with exponential cutoff, ${dN\over dE}\propto E^{-\Gamma}e^{-E/E_{\rm cut}}$).

The angle-integrated emission power corresponds to efficiency $\epsilon \approx 80$\%. The spectrum is close to PLEC with $\Gamma =0.9$ (the PLEC fit to synchrotron emission gives $\Gamma {\rm synch}=0.73$, but the distribution over critical energies $E_{\rm c}$ flattens it). 

The cutoff photon energy for the angle-integrated emission is, in physical units,
\be \label{cut}
E_{\rm cut}\approx 1.9 {mc^2\over \alpha}{\rm Ar}^{3/8},
\ee
where $mc^2=0.511$MeV is the electron mass, $\alpha ={e^2\over c\hbar}={1\over 137}$ is the fine structure constant, and we have introduced the {\it Aristotle number} of a pulsar ${\rm Ar}$,
\be \label{ari}
{\rm Ar}\equiv {L_{\rm sd}\over L_e}\left( {R_{\rm lc}\over r_e}\right) ^{-2/3},
\ee
where $r_e={e^2\over mc^2}=2.8\times 10^{-13}$cm is the classical electron radius and $L_e={mc^3\over r_e}=8.7\times 10^{16}$erg/s is the classical electron luminosity. As we show in \S\ref{deri}, AE is applicable to pulsars with ${\rm Ar}\gg 1$, which is true for all Fermi pulsars. In astrophysical notation, eq.(\ref{cut}) reads
\be \label{cuta}
E_{\rm cut}\approx 5.2L_{34}^{3/8}P_{\rm ms}^{-1/4}{\rm GeV},
\ee
which agrees with the crude estimate of \cite{Gruzinov2012}. 

\begin{figure}[bth]
  \centering
  \includegraphics[width=0.48\textwidth]{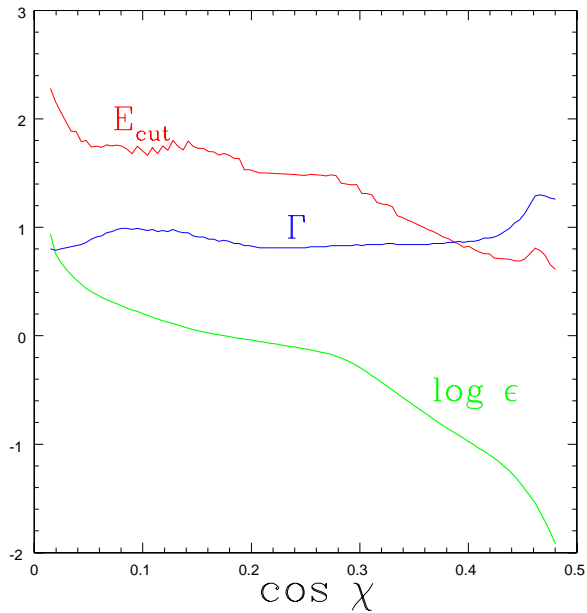}
\caption{Observed photon cutoff energy $E_{\rm cut}$, photon index $\Gamma$, and efficiency $\epsilon$ vs. the observation angle $\chi$.} \label{obs}
\end{figure}

Observed properties of emission are obtained by counting only photons emitted into a fixed infinitesimal solid angle. Let $\chi$ be the observation angle: the angle between the spin axis and the direction to the observer. Then we select only photons emitted by particles with $|v_z|\approx \cos \chi$. We get the observed efficiency $\epsilon$, photon index $\Gamma$, and photon cutoff energy $E_{\rm cut}$ shown in Fig.(\ref{obs}). 

Serious comparison with observations is only possible after a 3D simulation. But Table \ref{comp} looks promising.

\begin{table}[bth]
\centering
\begin{tabular}{ccccccc}

PSR & $P_{\rm ms}$ & $L_{34}$ & $E_{\rm cut}$ & $E_{\rm cut}$, (\ref{cuta}) & $\Gamma$ & $\epsilon$, \% \\

\\

0106 & 83 & 2.9 & $2.7\pm 0.6$ & 2.6 & $1.2\pm 0.2$ & $71^{+60}_{-31}$ \\

0357 & 444 & 0.6 & $0.8\pm 0.1$ & 0.9 & $1.0\pm 0.1$ & ... \\

0622 & 333 & 2.7 & $0.6\pm 0.1$ & 1.8 & $0.6\pm 0.4$ & ... \\

1057 & 197 & 3.0 & $1.4\pm 0.1$ & 2.1 & $1.0\pm 0.1$ & $14\pm 10$ \\

1741 & 414 & 0.9 & $0.9\pm 0.1$ & 1.1 & $1.1\pm 0.1$ &  $22\pm 3$\\

1836 & 173 & 1.1 & $2.0\pm 0.1$ & 1.5 & $1.2\pm 0.1$ & $180^{+200}_{-100}$ \\

1957 & 375 & 0.5 & $1.0\pm 0.2$ & 0.9 & $1.3\pm 0.2$ & ... \\

2030+4 & 227 & 2.2 & $1.7\pm 0.3$ & 1.8 & $1.6\pm 0.1$ & ... \\

2055 & 320 & 0.5 & $1.1\pm 0.1$ & 0.9 & $1.0\pm 0.1$ & ... \\

2139 & 283 & 0.3 & $1.3\pm 0.3$ & 0.8 & $1.3\pm 0.2$ & ... \\

\end{tabular}
\caption{\label{comp} All young Fermi pulsars with spin-down power $L_{34}<3$, from \cite{Fermi2013}, $E_{\rm cut}$ is in GeV. The fifth column calculates $E_{\rm cut}$ from eq.(\ref{cuta}).}
\end{table}

\section{AE and weak pulsar}\label{deri}
Here we derive the equations used in \S\S\ref{magnetos},\ref{emiti}. We first derive AE and calculate radiation in AE. Then we show that the magnetosphere calculated in \S\ref{magnetos} is realizable as an AE flow emanating from the star. 

\subsection{AE, radiation in AE}

For a generic electromagnetic field, a one-parameter family of Lorentz frames exists at any event, such that ${\bf E}$ is parallel to ${\bf B}$ in these frames. Assume that in these frames a positive charge moves at the speed of light along ${\bf E}$ and a negative charge moves at the speed of light along $-{\bf E}$. Written in an arbitrary Lorentz frame, this gives the basic AE equation (\ref{vel}).

Of course, the charges actually move slower than light. The terminal Lorentz factor $\gamma$ is reached when the curvature radiation power balances the accelerating power of $E_0$:
\be 
\pm ec{\bf v}_\pm \cdot {\bf E}=ecE_0={2\over 3}e^2cK^2\gamma ^4,
\ee
where the curvature of the trajectory $K$ can be calculated using the approximate speed of light motion given by eq.(\ref{vel}). Knowing the terminal Lorentz factor $\gamma$, one gets the critical energy of the emitted photon spectrum (\ref{ecrit}).

The distance over which a charge needs to travel in order to get accelerated to the terminal Lorentz factor $\gamma$ is $\sim {\gamma mc^2\over eE_0}$. Demanding that this distance be much smaller than $R_{\rm lc}$, estimating the curvature as $K\sim R_{\rm lc}^{-1}$, and estimating the electric field from $L_{\rm sd}\sim cE_0^2R_{\rm lc}^2$, we get the AE applicability condition 
\be
{\rm Ar}\equiv {L_{\rm sd}\over L_e}\left( {R_{\rm lc}\over r_e}\right) ^{-2/3}\gg 1.
\ee
With the same estimates, the critical photon energy given by (\ref{ecrit}) is
\be 
E_{\rm c}\sim {mc^2\over \alpha}{\rm Ar}^{3/8},
\ee
in agreement with the numerical result (\ref{cut}).

\subsection{AE magnetosphere}

In brief, the recipe of \S\S\ref{magnetos},\ref{emiti} works because plasma multiplicity in the radiation zone is zero, meaning that only a single charge species is present at any point. This allows to infer the particle density from the electric charge density and calculate the radiation. Zero multiplicity also means that the Ohm's law (\ref{ohm}) is exact in the radiation zone. In the force-free zone eq.(\ref{ohm}) is wrong, but the actual form of the Ohm's law in the force-free zone is irrelevant. 

Consider what happens in real weak pulsars (as opposed to what happens in the calculation of \S\ref{magnetos}). Near the star, the standard avalanche mechanism \cite{Ruderman1975} keeps working to (nearly) nullify $E_0$. This requires permanent pair production in the near-star zone; and some of the charges then flow out, thus bringing the electric charge and electric current to the space around the star. Let $\rho_\pm$ be the $e$-normalized number densities of positrons and electrons, so that the electric charge density is $\rho=\rho_+-\rho_-$. The electric current density is then 
\be\label{ohm1}
    {\bf j}={\rho {\bf E}\times {\bf B}+P(B_0{\bf B}+E_0{\bf E})\over B^2+E_0^2},
\ee
where $P$ ('Rho') is the total particle density
\be\label{P}
    P=\rho_++\rho_-.
\ee

{\it Iff} the plasma multiplicity is zero, the Ohm's law of \S\ref{magnetos}, eq.(\ref{ohm}), coincides with the true Ohm's law, eq. (\ref{ohm1}). But (though zero-multiplicity AE flows are theoretically possible, say Michel's monopole \cite{Michel1973}) the pulsar flow must have non-zero multiplicity. This is clear from the following argument. 

As we know (more precisely, will know after finishing this section), the star is surrounded by a force-free zone. In the force-free zone the density can be calculated from the pulsar magnetosphere equation \cite{Scharlemant1973}:  one can show that $\phi=\psi$, $A=A(\psi)$, and 
\be\label{fff}
(1-r^2)\rho = -{2\over r}\partial _r\psi +A{dA\over d\psi }.
\ee
Well inside the light cylinder, but not too close to the star, at $R_s\ll R\ll R_{\rm lc}$, the field is close to a pure dipole:
\be\label{dip}
\psi={r^2\over (r^2+z^2)^{3/2}};
\ee
the last term in the r.h.s. of eq.(\ref{fff}) can be ignored, and we get at high altitudes, $|z|\gg r$:
\be
\rho =-{4\over |z|^3};
\ee
i.e., at $R\ll R_{\rm lc}$, the charge density is negative in the entire current carrying region (for $R\ll R_{\rm lc}$, the corotation zone reaches high altitudes). Yet the current here has to flow both in and out of the star. In the negatively charged region, the only way to arrange for an out-going current using only out-going charges is to have both species present, with electrons moving out slower than positrons. 

\begin{figure}[bth]
  \centering
  \includegraphics[width=0.48\textwidth]{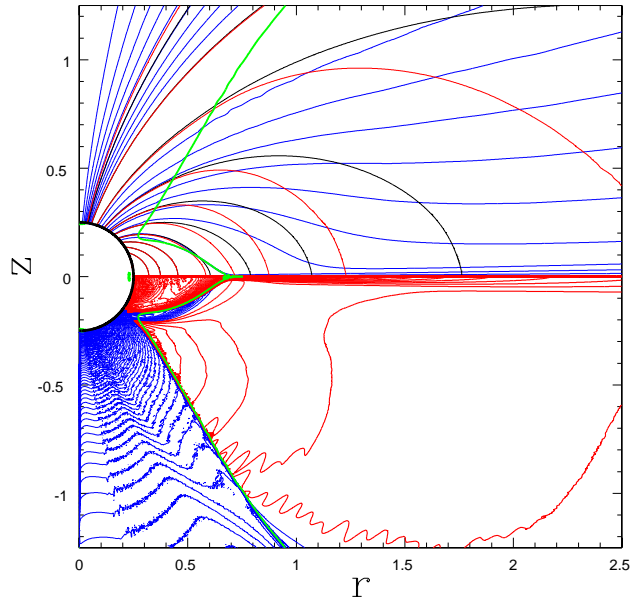}
\caption{Upper half -- same as Fig.(\ref{magn}). Lower half -- charge density $\rho$, $\pm 0.05*k^2$ with k=1,2,3,...; negative in thin blue, positive in thick red. } \label{dens1}
\end{figure}

Thus we do have current-carrying regions of non-zero multiplicity, and this means that we are knowingly using a wrong Ohm's law.  Nevertheless, we claim that the mean electromagnetic field, as given by $\psi$, $\phi$, and $A$ shown in the figures, is correct. This is because only the force-free zone has non-zero multiplicity, while in the radiation zone our flow is single-species and emanating from the force-free zone. So our Ohm's law is correct in the radiation zone.

In the force-free zone, the sign of $B_0$ fluctuates and the mean current is determined by the averaged value $\bar{B_0}$. In the real pulsar, in the out-going current part of the force-free zone: (i) $\bar{B_0}$ adjusts in such a way as to suspend electrons at  fixed poloidal positions (otherwise electrons would accumulate at the boundary between the force-free and radiation zones); (ii) with known $\bar{B_0}$, $\rho_+$ is determined by our correctly calculated poloidal current density (given by $A$); (iii) with known $\rho _+$, $\rho_-$ is determined by our correctly calculated $\rho$ (given by $\phi$). Our simulation is capable of modeling the real pulsar, because $\bar{B_0}_{\rm real}(\rho_++\rho_-)=\bar{B_0}_{\rm sim}|\rho |$.

The above analysis shows (once we prove that the flow of charge in the radiation zone emanates from the force-free boundary rather than from infinity, which we will do in a second) that our electromagnetic field is realizable as an AE flow emanating from the star. We must emphasize again that, at least in principle, many AE flows can exist,  and there is no way of knowing that the flow we have calculated is the one which is realized in nature. One must try to do an actual AE simulation. We did try, and to the extent that it worked, the AE simulations, both the density-field simulation and the particle simulation, agree with the results of \S\ref{magnetos}.

Now we must look at the charge density field $\rho$, Fig.(\ref{dens1}). If we ignore the thin blue tongue (electrons) just outside the corotation zone, then all is well. The high-altitude poloidal current is in-coming -- it is carried by the out-going electrons, both in the force-free and in the radiation zone (note a thin blue line near $r=1$, $z=-1.2$). The low-altitude poloidal current is out-going -- it is carried by the out-going positrons. In the out-going current part of the force-free zone the charge density is negative due to poloidally suspended electrons (meaning, as described above, that there exists a $\bar{B_0}$ which can do this). In the region of out-going electrons we can also have some suspended positrons (on the field lines which ultimately enter the radiation zone) or out-going positrons (on the field lines which forever remain in the force-free zone); the latter part of the flow can have arbitrarily high multiplicity, whose actual value is perhaps regulated by the near-star avalanche. 

\begin{figure}[bth]
  \centering
  \includegraphics[width=0.48\textwidth]{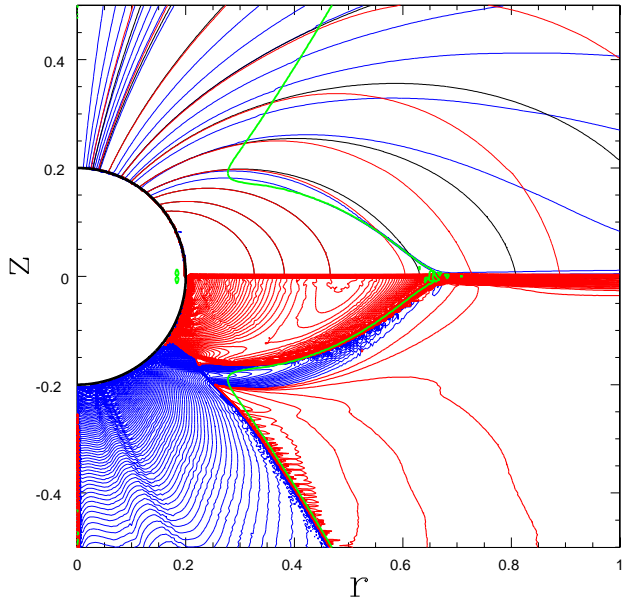}
\caption{Same as Fig.(\ref{dens1}), with $A_{\rm max}=0.30$. } \label{dens2}
\end{figure}

\begin{figure}[bth]
  \centering
  \includegraphics[width=0.48\textwidth]{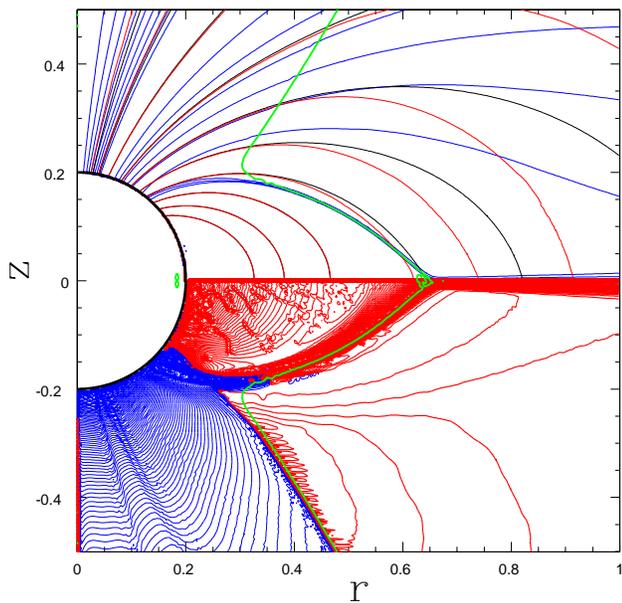}
\caption{Same as Fig.(\ref{dens1}), with $A_{\rm max}=0.32$. Non-zero multiplicity in the co-rotating zone (multiplicity parameter $f=0.7$ in eq.(\ref{P1})} \label{dens3}
\end{figure}

Now for the electron tongue. Formally, it is fatal for our attempt to interpret the wrong Ohm's law solution in terms of the true AE flow, for it requires that the electrons (flowing here towards the star) be created at the boundary of the corotation zone. However the tongue is thin and physically it clearly corresponds to a singular current flowing along (rather than emanating from) the boundary of the corotation zone. It certainly appears to be just a numerical issue. 

To confirm this, we performed a separate calculation with a smaller star, $R_s=0.2$, in a smaller, 2x4, box. First we  repeated the original calculation and got the same electron tongue Fig.(\ref{dens2}). Then we changed the Ohm's law in the corotation zone from eq.(\ref{ohm}) to eq.(\ref{ohm1}) with
\be \label{P1}
P=|\rho |+f{\sqrt{E^2+B^2}\over R},
\ee
where $R$ is the spherical radius and we call $f$ the multiplicity parameter; $f\sim 1$ has the potential of strongly affecting the resulting electromagnetic field. 

Because in the corotation zone both $E_0=0$ and $\bar{B_0}=0$, we are allowed to use any $P$ in the Ohm's law (\ref{ohm1}). Supposedly it models the same physics, but regularizes the singular current in a different way. We now get a fully satisfactory charge density field,  Fig.(\ref{dens3}), and, importantly, without significant changes in the calculated fields outside the tongue. 

As the tongue contributes but little to the calculated radiation (we checked this), we have decided to keep the calculation recipe of \S\ref{magnetos} simple and ignore the issue of proper regularization of the corotation zone. Obviously, when doing the 3D case, one will have to check that our simple recipe remains applicable. 

\section{Conclusion}

Barring unpleasant surprises from the future full AE calculations:

\begin{itemize}

\item We have calculated the emission spectrum of a weak axisymmetric pulsar from first principles.

\item Repeated in 3D, the same calculation is expected to give energy-resolved light curves which will agree with observations without any adjustable parameters.

\item To calculate non-weak pulsars, i.e. pulsars with non-negligible pair production near the light cylinder, a full AE code with explicit electron and positron densities is necessary (but obviously not sufficient: pair production and therefore also the photon kinetics, both X and gamma, must be explicitly treated).

\end{itemize}

\end{document}